\documentclass[reprint,superscriptaddress,amsmath,amssymb,aps,pra]{revtex4-2}

\usepackage{graphicx}
\usepackage{dcolumn}
\usepackage{bm}
\usepackage{float}
\usepackage{xcolor}
\usepackage{multirow}
\usepackage[colorlinks=true, allcolors=blue]{hyperref}
\usepackage{flafter}
\def\fnum@figure{\figurename\thefigure}
\renewcommand{\figurename}{Fig.}

\begin{document}

\preprint{APS/123-QED}

\title{Surface Plasmon Mediated Giant Goos–Hänchen and Imbert–Fedorov Shifts on a Corrugated Metal Surface}

\author{Arani Maiti}
\affiliation{Department of Physical Sciences, IISER-Kolkata, Mohanpur 741246, India}

\author{Sauvik Roy}
\email{sauvikroy3388@gmail.com}
\affiliation{Department of Physical Sciences, IISER-Kolkata, Mohanpur 741246, India}

\author{Abhi Mondal}
\affiliation{Department of Physical Sciences, IISER-Kolkata, Mohanpur 741246, India}

\author{Ayan Banerjee}
\affiliation{Department of Physical Sciences, IISER-Kolkata, Mohanpur 741246, India}

\author{Nirmalya Ghosh}
\affiliation{Department of Physical Sciences, IISER-Kolkata, Mohanpur 741246, India}

\author{Subhasish Dutta Gupta}
\email{sdghyderabad@gmail.com}
\affiliation{Department of Physical Sciences, IISER-Kolkata, Mohanpur 741246, India}
\affiliation{Tata Institute of Fundamental Research, Hyderabad, Telangana 500046, India}

\date{\today}

\begin{abstract}
    Enhanced beam shifts mediated by surface plasmon resonance (SPR) at metal–dielectric interfaces have been widely investigated. However, research on the associated Imbert-Fedorov or spin Hall shifts, driven by the spin–orbit interaction of structured light in structured interfaces, has been comparatively scarce and limited. We explore the reflection characteristics of generic polarized, non-paraxial light beams from a corrugated silver (Ag) interface, since surface corrugation can naturally couple the incident radiation modes to the surface excitations. In the vicinity of SPR, we report a significant enhancement in the beam shifts, attributed to the rapid variation of the specular reflection coefficient near its minima, resulting in amplified weak values. By carefully selecting the incident and projected polarization states of the beam, we achieve a pronounced spatial spin Hall effect. We also investigate vortex-induced beam shifts within this resonant regime, revealing distinctive signatures of the beam’s angular momentum. Furthermore, a comprehensive analysis is also presented for the conical diffraction geometry, wherein polarization conversions between \textit{p} and \textit{s} states are fully incorporated. Our work establishes the interplay of the spin-orbit interaction of light and the weak measurement approach as an important methodology in amplifying SPR effects, which may have important connotations in applications involving light at nanoscales.
\end{abstract}

\maketitle


\section{Introduction}

The conventional laws of reflection and transmission —- derived from elementary wave optics and based on a single plane-wave description -— prove inadequate for accurately characterizing the behavior of a coherent, polarized, and structured beam at an interface \cite{r1,r4}. The subtle deviations of the reflected and transmitted beams from predictions of standard wave optics have inspired extensive investigation into the underlying mechanisms, particularly the lateral Goos–Hänchen (GH) and transverse Imbert–Fedorov (IF) beam shifts across diverse optical platforms \cite{Artman,r1,r4}. A beam with finite cross-section can be regarded as a superposition of plane waves, and the GH and IF shifts arise from the spread of the plane waves (or the spread of wavevectors in $k$-space), the polarization states of individual plane-wave components, and their interactions with the optical system. GH and IF shifts have been studied in a variety of optical systems, including dielectric–metal interfaces \cite{Merano:07,LEUNG2007206}, multilayer thin films and coatings \cite{r1,r4,Sonipramana,golla2011pramana,KUMARIoptcomm,critical_coupling_beam,cpa_roy}, metasurfaces \cite{yallapragada2016observation,wang2021shifting}, anisotropic and birefringent media \cite{Qin:10}, photonic crystals \cite{PhysRevLett.108.123901}, and negative-index materials \cite{GHlefthanded}. Recently, these shifts have found practical applications in areas such as precision metrology \cite{s19092088,sensorJPCA,sensorBiosensors}, refractive-index sensing \cite{KUMARIoptcomm}, thickness and temperature characterization  \cite{temperature}, layer counting \cite{layer}, and biosensing \cite{karimi2025development}. 

Since the physics of beam shifts is fundamentally governed by the complex Fresnel reflection and transmission coefficients, it becomes naturally compelling to explore the interaction of structured light in structured plasmonic interfaces \cite{PhysRevBlocalsurface}. In metals, surface plasmon resonance (SPR) is highly sensitive to the excitation momentum, causing specific $k$- vectors within the beam spectrum to be strongly attenuated while others remain largely unaffected. This selective interaction modifies the spatial distribution of the reflected field, leading to a pronounced enhancement of the beam shift near SPR. Furthermore, the surface texture can give rise to field enhancements across various diffraction orders, where complex polarization conversion and the mixing of orthogonal polarization components occur as a consequence of the spin–orbit interaction (SOI) of light. On another note, the weak measurement technique - well known in the context of quantum systems \cite{PhysRevD} - has been extensively used in optics \cite{weak_science,weak_IOP,Dennis_weak_IOP,weak_optics_letters, Realization_of_a_measurement_of_a_weak_value,Roadmap_on_Weak_Measurements_in_Optics,Bliokh2016} to significantly enhance low intensity optical signals by a judicious choice of polarization states of light during the measurement process. It could therefore be expected that even the effects of SOI could be similarly amplified by coupling them with a weak measurement scheme.
\par
In this paper, we explore this interesting domain of research, where we consider a corrugated silver (Ag) surface with a sinusoidal height modulation \cite{r5.1,book1,r6}, illuminated by polarized Gaussian and Laguerre–Gaussian (LG) beams—without and with orbital angular momentum (OAM), respectively. The one-dimensional sinusoidal grating represents one of the simplest surface textures with significant potential for manipulating light through spin–orbit interaction (SOI). Other forms of surface structuring have been explored to break reflection symmetry and introduce additional geometric phase \cite{ErezHasman,PhysRevLett.119.253901,PatriceGenevet}, enabling phase engineering for beam steering \cite{yu2011light} and providing the necessary momentum matching for the excitation of surface modes. Various other configurations have also been investigated to achieve ultra-high-Q resonances \cite{PhysRevLett.119.253901,book1,highQ} and to realize bound states in the continuum (BICs) \cite{bicgrating}. We have implemented an improved form of the generalized angular spectrum method, originally developed by Bliokh and Aiello \cite{r1}, which enables accurate modeling of structured beams with finite spatial extent, modal structure, and polarization. The illumination configuration encompasses both the in-plane diffraction geometry—where the central $\mathbf{k_c}$-vector of the incident beam and the grating vector lie in the same plane ($\Phi=0$) —and the conical diffraction geometry, in which they are non-coplanar ($\Phi \neq 0$). The pronounced variation of the specularly reflected mode arising from surface plasmon resonance (SPR) is examined in the zeroth-order (radiation mode) and first-order (surface mode) diffracted components. Importantly, the beam shift phenomena (reflected beam profiles) are observed exclusively in the radiation mode. It is noteworthy that, owing to the weak spin–orbit interaction at an interface mediated by the Fresnel reflection coefficients, the off-axis (non-central) $k$-vectors in the beam spectrum undergo almost identical changes in complex amplitude as the central wavevector $\mathbf{k_c}$. Consequently, the transversely polarized scalar components (pointers) undergo small spatial and angular displacements -- commonly referred to as beam shifts—- which can be interpreted as complex quantum weak values. This correspondence forms the basis for the weak measurement of the polarization state of the reflected beam \cite{weak_science,weak_IOP,Dennis_weak_IOP,weak_optics_letters, Realization_of_a_measurement_of_a_weak_value,Roadmap_on_Weak_Measurements_in_Optics,Bliokh2016}. During interaction with the interface, the initially homogeneous polarization of the incident beam becomes modified, and these beam shifts can be significantly amplified by post-selecting the reflected beam in a nearly orthogonal polarization state. Here, we present a comprehensive theoretical analysis and numerical simulations exploring the extent to which the Goos–Hänchen (GH) and Imbert–Fedorov (IF) beam shifts can be enhanced through post-selection of the diffracted orders into various polarization states. 

The structure of this paper is organized as follows. Section \ref{sec:theory} introduces the theoretical framework used to calculate the field components corresponding to various diffraction orders. Section \ref{sec:results} presents the numerical simulations, followed by a detailed analysis and discussion of the results. Finally, Section \ref{sec:conclusion} provides a concise summary of the key findings and concluding remarks.

\section{Theoretical framework and mathematical formulation}\label{sec:theory}
\begin{figure}[ht!]
    \centering
    \includegraphics[width=\linewidth]{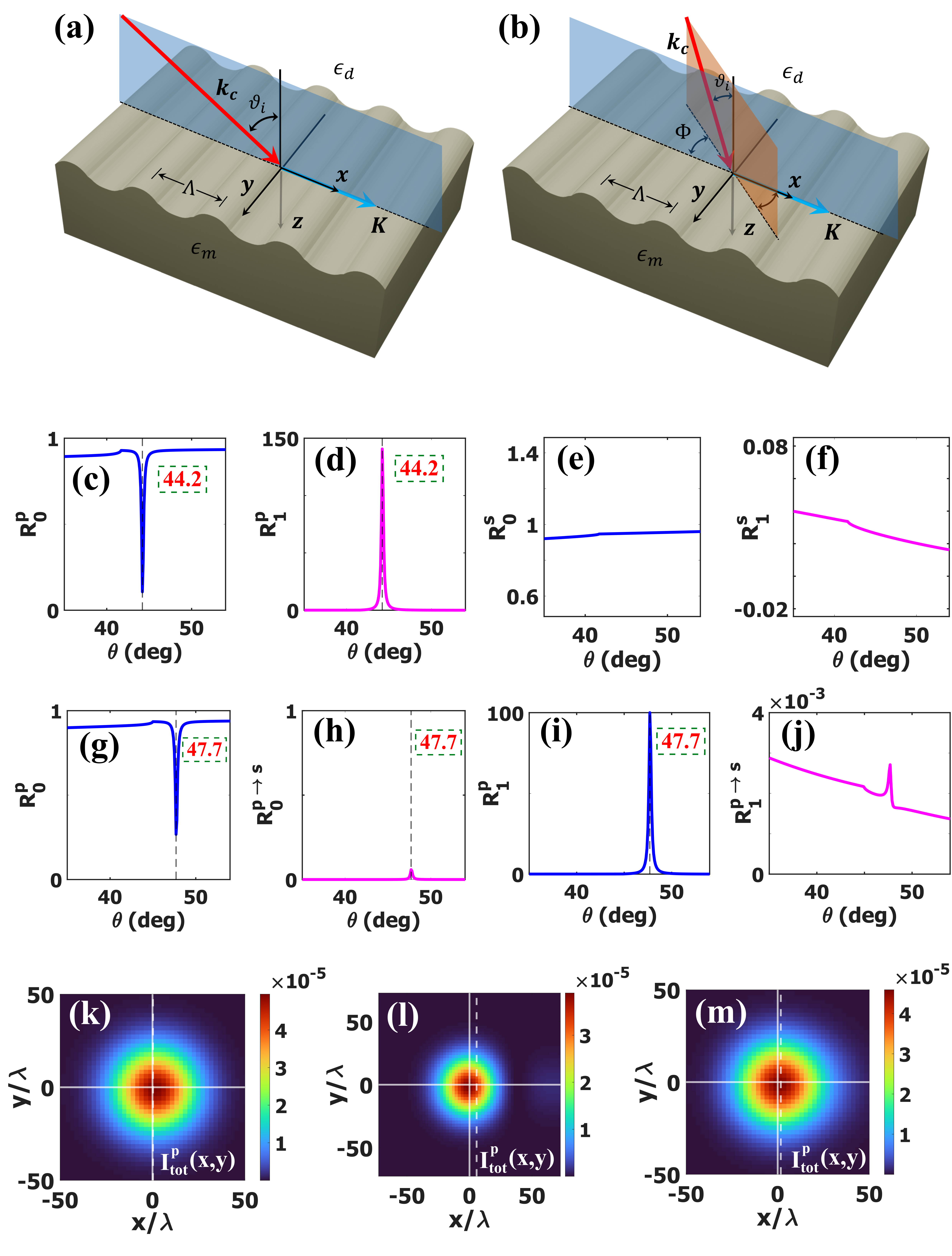}
    \caption{Schematics showing (a) in-plane and (b) conical diffraction on a corrugated metallic surface, where the grating vector $\mathbf{K}$ (blue arrow) lies within or outside the plane of incidence, respectively. Only the central wave vector $\mathbf{k_c}$ (red arrow) of the incident beam spectrum is shown, oriented at an azimuthal angle $\Phi$ and an angle of incidence $\vartheta_i$. Angular dependence of (c, e) zero-order and (d, f) +1-order reflected intensities for \textit{p}- and \textit{s}-polarized plane-wave illuminations for the in-plane diffraction geometry. (g–j) Corresponding intensities for the conical diffraction with an azimuthal angle $\Phi=35^{\circ}$ for an incident $p$ polarized plane wave. Giant spatial GH shifts near the SPR angle for a \textit{p}-polarized incident Gaussian beam at three different incident angles: (k) $41.75^{\circ}$, (l) $44.2^{\circ}$, and (m) $45.5^{\circ}$, respectively. Dashed (solid) crosshair represents the reflected (incident) beam centroid.}
    \label{fig1}
\end{figure}

Consider a corrugated metal-dielectric interface with the surface profile given by (Figs. \ref{fig1} (a),(b))
\begin{equation}\label{eq1}
	z=a\sin{Kx},
\end{equation}
illuminated by a structured beam with its central wave vector at an angle of incidence $\vartheta_i$. Further the beam is assumed to be incident on the mean surface ($z=0$) at its waist. In Eq. \ref{eq1}, $a$ is the grating amplitude and $K=2\pi/\Lambda$ is the grating vector. The metal is characterized by a complex permittivity $\epsilon_m $, while the dielectric (in our case, air) permittivity is given by $\epsilon_d$.
\par
Using the formalism developed in Ref.~\cite{r5.1}, one can calculate all the complex amplitudes for the propagating and evanescent diffraction orders for the reflected light for any plane wave component of the incident beam. We mostly focus on specular order reflected beam profiles. Next, we generalize the plane wave results to a beam using the angular spectrum formalism originally developed by Bliokh and Aiello~\cite{r1} to investigate beam shift phenomena associated with the reflection of structured beams. To extend the framework beyond the constraints of the strict paraxial approximation, we refine the method by incorporating a broader transverse momentum distribution, thereby capturing essential non-paraxial effects with greater accuracy~\cite{r2,r3}. Initially, we used a Gaussian beam propagating along the z-axis (in the beam frame) with~\cite{r1} 
 \begin{equation}
\left|\mathbf{E}_i\right\rangle=\frac{w_0}{\sqrt{2 \pi}} \exp \left(-\left(k_x^2+k_y^2\right) w_0^2 / 4\right)\left(A_p \mathbf{e}_p+A_s \mathbf{e}_s\right).
\label{eq2}
\end{equation}
Here, the second parenthesis specifies the polarization state. The polar and azimuthal angles of the off-axis wave vectors can be given by the following expressions~\cite{r4}.
\begin{equation}
\theta_i  =\tan ^{-1}\left(\frac{\sqrt{k_y^2+\left(k_x \cos \vartheta_i+k_z \sin \vartheta_i\right)^2}}{-k_x \sin \vartheta_i+k_z \cos \vartheta_i}\right),
\label{eq3}
\end{equation}
\begin{equation}
\phi_i  =\tan ^{-1} \frac{k_y}{k_x \cos \vartheta_i+k_z \sin \vartheta_i},
\label{eq4}
\end{equation}
where $k_z=\sqrt{k_0^2-\left(k_x^2+k_y^2\right)}$. Subsequently, we also study the beam shifts for the LG beam given by the expression ~\cite{r1}
\begin{equation}
\begin{aligned}
\left|\mathbf{E}_i\right\rangle \propto & \frac{w_0}{\sqrt{2 \pi}} \exp \left(-\left(k_0 w_0\right)^2 \theta_z^2 / 4\right) \theta_z^{|\ell|} e^{i \ell \phi+i k_0\left(1-\theta_z^2 / 2\right) z} \\
& \times\left(A_p \mathbf{e}_p+A_s \mathbf{e}_s\right),
\label{eq5}
\end{aligned}
\end{equation}
where $\theta_z=\sqrt{k_x^2+k_y^2} / k_0, \phi=\tan ^{-1}\left(k_y / k_x\right)$ and vortex charge $\ell= 0, \pm1,\pm2,\dots$. Throughout the analysis, the incident field is taken to be monochromatic, varying in time as $\exp(- i\omega t)$.
By applying the Fresnel reflection coefficients and the appropriate rotation matrices for each individual wave vector component of the incident beam, we numerically evaluate the reflected field, enabling a complete description of the angular response of the beam. So, the output spectra are given by\cite{r1,r2}:
\begin{equation}
|\mathbf{E}_r\rangle = U_r^{\dagger} F_r U_i |\mathbf{E}_i\rangle,
\label{eq6}
\end{equation}
where
\begin{equation}
U_r = \hat{R}_y(\theta_r) \hat{R}_z(\phi_r) \hat{R}_y(-\vartheta_r),
\label{7}
\end{equation}
and 
\begin{equation}
F_r=
\begin{pmatrix}
     r_p  & 0 \\
        0 & r_s 
    \end{pmatrix}
    \label{eq8}
\end{equation}
Here, the subscripts $ i $ and $ r $ represent the quantities associated with the incident and reflected fields, respectively. The matrices $ R_x $, $ R_y $, and $ R_z $ denote rotations about the $ x $-, $ y $-, and $ z $-axes in the laboratory coordinate system. After performing the inverse Fourier transformation of the equation~(\ref{eq6}), we compute the corresponding spatial field components, and the determine the GH and IF shifts by calculating the displacement of the centroid \cite{r2,critical_coupling_beam,cpa_roy} of the reflected intensity distributions through spatial averaging.     
\section{Results and discussion}\label{sec:results}
\subsection{SPR mediated enhancement of Goos–Hänchen shift at corrugated metal interface in in-plane configuration}

The method described above is implemented in MATLAB, and the following sections present the results of our numerical calculations. For our simulation, we have used the following material and beam parameters: wavelength $ \lambda = 0.6328 \mu m $, $ \epsilon_m = -16.3 + i 0.53 $, $\epsilon_d=1$, grating period $ \Lambda = 1.888 \mu m $ and $ a = 0.025 \mu m $, beam waist $w_0 =15\lambda$. To start with, we examine the reflected orders for a single plane wave incidence rather than the Gaussian beam itself in the case of in-plane diffraction geometry~\cite{r5.1,r5,book1}. In this geometry, the central wavevector $\mathbf{k_c}$ and the grating vector $\mathbf{K}$ lie in the same plane (Fig.~\ref{fig1}(a)). Note that surface plasmon resonance (SPR) can be excited only by a $p$-polarized incident beam (Figs.~\ref{fig1}(c) and ~\ref{fig1}(d)), owing to the presence of a longitudinal electric field component along the grating vector. A pronounced field enhancement is observed (Fig.~\ref{fig1}(d)) in the $ +1 $ evanescent order near the SPR excitation angle of $ 44.2^\circ~(\theta_p)$ together with a sharp dip in the $0^{\text{th}}$ order radiation mode (Fig. \ref{fig1} (c)) for \textit{p}-polarized light, which is attributed to the excitation of surface plasmons. As a result of these sharp variations of the reflection coefficients $r_p$, the quantity $ \frac{\partial r_p}{\partial \vartheta_i} / r_p $, which is proportional to the Goos–Hänchen (GH) shift~\cite{r1}, increases dramatically. Consequently, a substantial spatial GH shift is expected near the surface plasmon resonance (SPR) angle for the reflected specular Gaussian beam. In contrast, a \textit{s}-polarized incident light fails to excite surface plasmons (Figs.~\ref{fig1}(e) and ~\ref{fig1}(f)) in this geometry.
\par
The spatial intensity profiles of the reflected beams for the incident Gaussian beam are shown in Figs.~~\ref{fig1}(k),~\ref{fig1}(l), and ~\ref{fig1}(m) for three different angles of incidence $41.75^\circ$ (before resonance), $44.2^\circ$ (at resonance), and $45.5^\circ$ (after resonance) (Fig. \ref{fig1} (k), (l), (m)). The corresponding GH shifts are found to be $0.2534\lambda$ , $5.6967\lambda$, and $1.7091\lambda$, respectively. It is clear that as the angle of incidence approaches the surface plasmon resonance (SPR) condition, the spatial Goos–Hänchen (GH) shift becomes significantly enhanced with the maximum $5.6967\lambda$ at resonance. Also, the reflected beam exhibits a marked asymmetry due to SPR-induced distortion. At the resonance angle, the spatial harmonics with k-vectors responsible for the SPR excitation get absorbed, resulting in a split spectrum (not shown), and the left and right lobes of the spectra can interfere destructively in the space domain, resulting in the main spot along with a faint spot to the right. Thus, there is a splitting in the reflected beam profile with a faint side lobe, which has earlier been discussed in Ref \cite{r2}. In contrast, \textit{s}-polarized incident Gaussian beam can not lead to similar effects in in-plane configuration since there is no surface excitation. 

The conical diffraction geometry~\cite{r5,r6}, illustrated in Fig.~\ref{fig1}(b), corresponds to a scenario where the grating vector is at an azimuthal angle $\Phi$ with respect to the plane of incidence. In this case, for an incident $p$-polarized light, partial conversion from $p$ to $s$-polarization is observed (Fig. \ref{fig1} (h)). However, the resonance occurred is not strong (Fig. \ref{fig1} (j)). This polarization conversion is discussed more elaborately in sec \ref{sec:conical}. As expected, the $p$-polarized beam excites a plasmon resonance at a slightly higher angle of incidence of $47.7^{\circ}$ compared to $44.2^{\circ}$ for the in-plane geometry (Fig. \ref{fig1} (g), (i)) because of the effective change in the grating period. Thus, both $s$- and $p$-polarized Gaussian beams can excite resonances at a corrugated metal–dielectric interface.

\par
\begin{figure}[ht!]
    \centering
    \includegraphics[width=\linewidth]{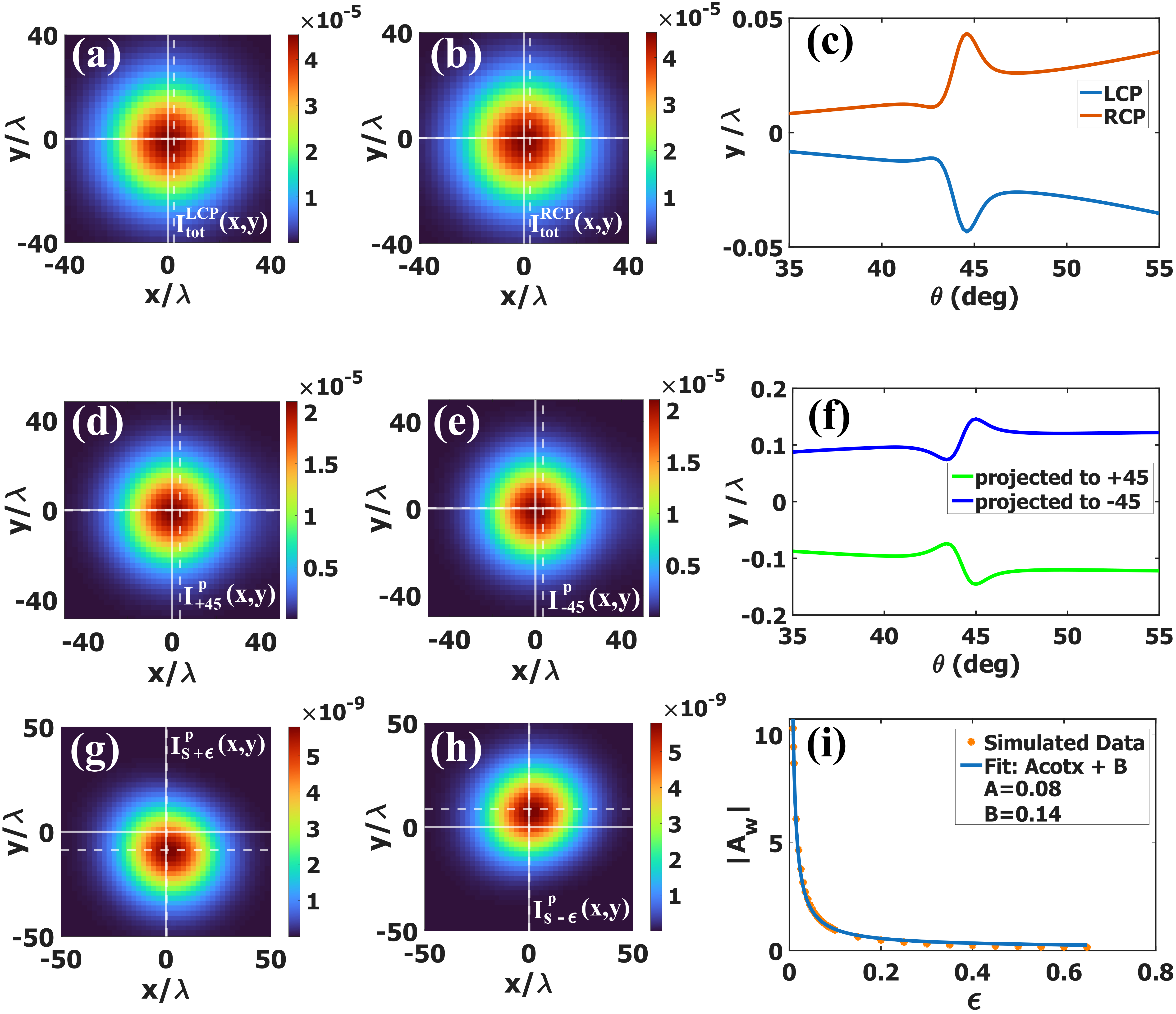}
    \caption{Reflected beam intensity profiles for incident (a) LCP and (b) RCP Gaussian beams at an angle of incidence $\vartheta_i = 44.6^\circ$. (c) Transverse IF shifts as a function of the angle of incidence for the incident LCP (blue line) and RCP (red line) beams. Reflected beam intensities for the (d) $+45^{\circ}$ ($I^{p}_{+45}$) and (e) $-45^{\circ}$ ($I^{p}_{-45}$) diagonal polarization projections correspond to an incident \textit{p}-polarized Gaussian beam at an angle of incidence $\vartheta_{i} = 45^{\circ}$. (f) IF shifts as a function of the angle of incidence for the $\pm45^{\circ}$ diagonally projected states. IF shifts of the reflected beams projected onto the nearly orthogonal states with (g) offset $\epsilon = +0.01$ and (h) offset $\epsilon = -0.01$. The incident beam is \textit{p}-polarized and impinges at an angle of incidence $\vartheta_{i} = 40^{\circ}$. (i) Magnitude of the weak value $|A_w|$ of the spatial IF shift as a function of offset $\epsilon$ for nearly orthogonal pre- and post-selected states. Here also, the dashed (solid) crosshair represents the reflected (incident) beam centroid.}
    \label{fig2}
\end{figure}
\subsection{Different strategies to enhance the spin Hall shift near SPR for in-plane configuration}
We now focus on the Imbert–Fedorov (IF) shifts ~\cite{r1} for incident circularly polarized Gaussian beams impinging at an angle near the SPR angle in the in-plane diffraction geometry. The characteristic magnitude of the spatial IF shift for an incident circularly polarized beam is given by~\cite{r1,weak_IOP}:
\begin{equation}
Y_{shift}\propto\pm\frac{i\cot \vartheta_i}{k_0}\frac{r_p^*r_p\left(1+\frac{r_s}{r_p}\right)+r_s^*r_s\left(1+\frac{r_p}{r_s}\right)}{\left|r_p\right|^2+\left|r_s\right|^2}
\label{eq9}
\end{equation}
The imaginary and real parts of the $Y_{shift}$ contribute to the spatial and angular IF shifts, respectively. The magnitudes of the IF shifts are found to be equal for the incident circularly polarized beams but in the opposite directions for opposite helicities. Specifically, $Y_{shift}= -0.0433\lambda$ for the LCP and $Y_{shift}=+0.0433\lambda$ for the RCP Gaussian beam for the angle of incidence $\vartheta_i=44.6^{\circ}$. In contrast, the spatial GH shifts are found to be $2.3202\lambda$ and in the same direction for both the LCP and RCP beams. Typically, the IF shifts near the SPR condition are much smaller for the incident circularly polarized beams as compared to the GH shifts for the plane polarized lights. However, an enhancement of the spatial IF shift is observed at an angle of incidence of $\vartheta_i = 44.6^\circ$ (Figs. ~\ref{fig2}(a) and ~\ref{fig2}(b)). The angular dependence of the transverse IF shifts for the incident LCP and RCP beams are shown in Fig.~~\ref{fig2}(c). It is to be noted that all the field profiles and the line plots pertaining to the shifts are generated using our improved method that goes beyond the strict paraxial approximation. Nevertheless, the shifts align well with the approximate analytical results in Eqs. \ref{eq9}--\ref{eq11}.

Next, we discuss two strategies to enhance the spin-Hall effect. The first involves a careful selection of the projected polarization state~\cite{weak_IOP} onto which the reflected beams need to be projected. Typically, the incident Gaussian beams are assumed to be uniformly polarized. However, due to the spread of wavevectors around the central wavevector $\mathbf{k}_c$, the plane of incidence varies for different spatial harmonics. As a result, the reflected beam, after interacting with the surface, is no longer uniformly polarized. This variation in polarization leads to distinctly different centroid shifts for various projected components of the reflected beam. We have chosen two orthogonal diagonal polarizations ($\pm45^{\circ}$) as the post-selection polarization state. The transverse IF shifts for the incident \textit{p}-polarized Gaussian beam, after filtering out the $\pm45^{\circ}$ components from the reflected beam, are given by~\cite{r1,weak_IOP}:
\begin{equation}
Y^p_{\pm45}\propto \pm \cot \vartheta_i\left(1+ \frac{r_s}{r_p} \right)
\label{eq10}
\end{equation}
It is to be noted that for the in-plane diffraction geometry and near the SPR angle, $r_p$ exhibits a sharp dip, while $r_s$ remains finite (Figs.~\ref{fig1}(c) and ~\ref{fig1}(e)). Because of the small value of $r_p$, Eq.~\ref{eq10} predicts an enhancement of the spatial IF shift near the SPR. These amplified beam shifts are captured in our simulations and are shown in Figs.~\ref{fig2}(d) and ~\ref{fig2}(e). The transverse spatial shifts are found to be $-0.1457\lambda$ and $+0.1457\lambda$ for the $\pm45^\circ$ projected reflected beams, respectively. The angular dependence of the IF shifts is shown in Fig.~\ref{fig2}(f). 

\par
The next strategy to achieve a giant spin Hall shift is based on the weak value amplification~\cite{Dennis_weak_IOP,Realization_of_a_measurement_of_a_weak_value,Roadmap_on_Weak_Measurements_in_Optics} protocol. As mentioned earlier, for the near-paraxial beam under consideration, the central wavevector is accompanied by a small spread of wavevectors around it. When such a beam is reflected, the reflection matrix acts differently on each of these plane-wave components. As a result, the polarization of the beam becomes weakly coupled to its spatial degrees of freedom, leading to a spatial beam shift that is analogous to the shift of a measurement pointer in weak measurement theory. In this approach, we first select a \textit{p}-polarized input state, represented by $(1,\,0)^{T}$, and then project the reflected beam onto a nearly orthogonal state $(\epsilon,\,1)^{T}$, where $\epsilon$ denotes a small offset parameter. In this case, the formula for the spin Hall shift gets modified as:
\begin{equation}
Y_{A_w}\propto \pm \frac{1}{\epsilon}\cot \vartheta_i\left(1+ \frac{r_s}{r_p} \right)
\label{eq11}
\end{equation}
By setting the offset parameter to $ \epsilon = \pm 0.01 $, a pronounced amplification of the spin Hall effect of light is observed, with the corresponding weak values of $ \mp 8.6756\,\lambda $, as illustrated in Figs.~\ref{fig2}(g) and \ref{fig2}(h).
As expected, these amplified spatial IF shifts are accompanied by reductions in the reflected beam intensities. Fig.~\ref{fig2}(i) presents the dependence of the weak value $|A_w|$ on the offset value $\epsilon$, which follows the characteristic $\cot(\epsilon)$ behavior (Eq. \ref{eq11}).
\par

\subsection{Vortex induced beam shifts}
\begin{figure}[ht!]
    \centering
    \includegraphics[width=\linewidth]{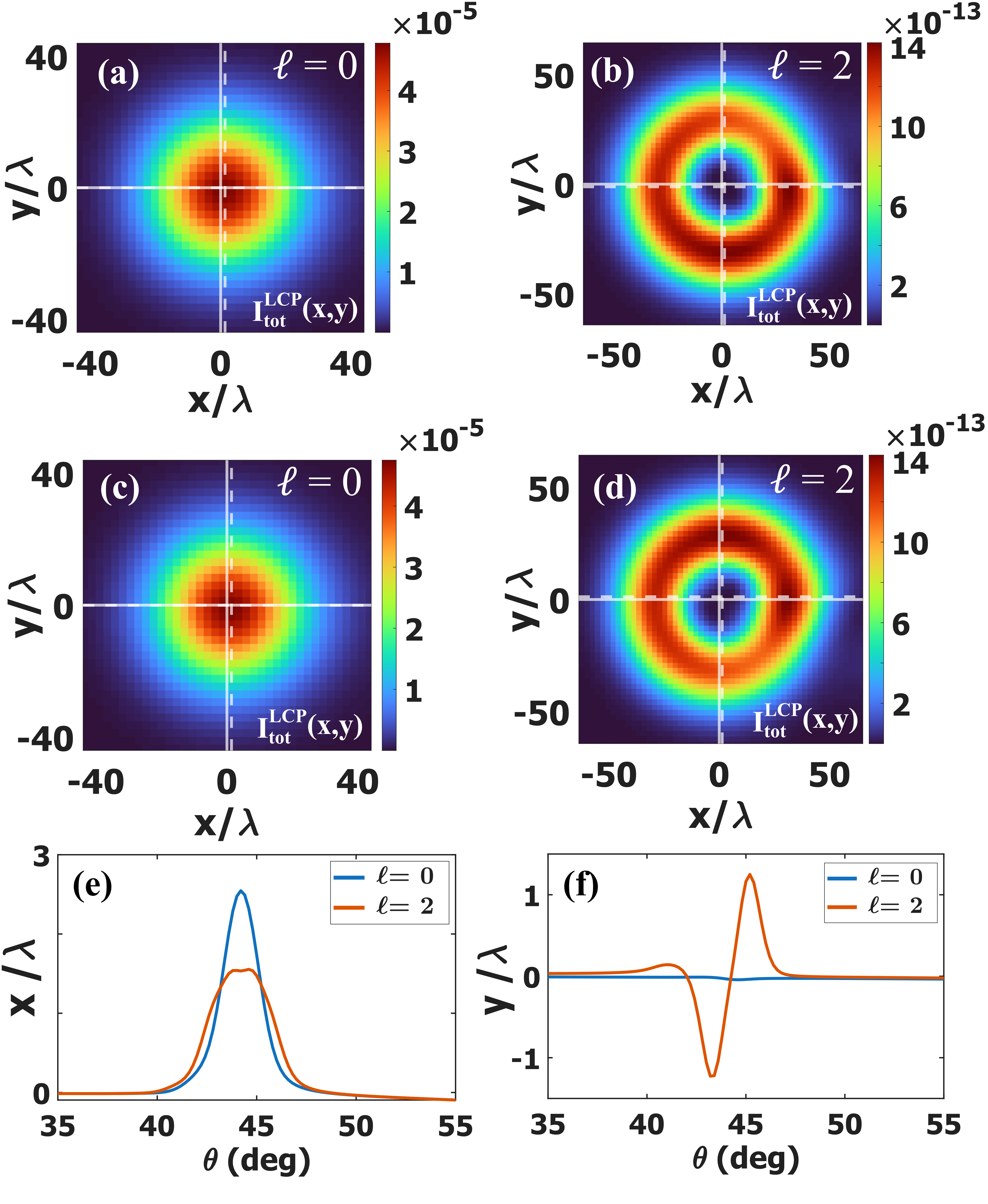}
    \caption{Intensity profiles of the reflected beams for the incident (a) LCP Gaussian beam ($\ell=0$) and (b) LCP LG beam ($\ell=2$) at an incidence angle of $43.2^{\circ}$ ($\theta<\theta_p$). The transverse IF shift for the LG beam is more pronounced compared to the Gaussian beam. (c,d) Similar intensity profiles of the reflected beams for an angle of incidence of $45.2^{\circ}$ ($\theta>\theta_p$). Variation of the spatial (e) GH and (f) IF shifts with the angle of incidence for the incident LCP Gaussian and LCP LG beams. As in Fig.~\ref{fig1}, the white dashed (solid) crosshair denotes the reflected (incident) beam centroid.}
    \label{fig3}
\end{figure}

LG beams -- since they carry orbital angular momentum -- exhibit richer effects as compared to Gaussian beams. Here, incident LG beams with a vortex charge of $\ell=2$ are examined for SOI signatures. 
Due to the complex vortex structure of the
beam, one can get additional shifts that are influenced by both the polarization state and the vortex charge. Spatial GH shift becomes coupled to the angular IF shift, while the spatial IF shift couples to the angular GH shift~\cite{r1,r8}, with vortex charge defining the coupling strength. Fig.~\ref{fig3}(a)--(d) presents the spatial GH and IF shifts for LCP Gaussian and LG beams, both before and after the SPR resonance condition. A detailed comparison of the shifts is given in Table \ref{tab:placeholder_label}.

\begin{table}[h]
    \centering
    \begin{tabular}{|c|c|c|c|}
        \hline
        \textbf{Angle} & \textbf{Vortex charge} & \textbf{GH Shift} & \textbf{IF Shift} \\ \hline
        $\theta < \theta_p$ & $\ell = 0$ & $+1.3530\lambda$ & $-0.0133\lambda$ \\ 
        $43.2^\circ$ & $\ell = 2$ & $+1.3350\lambda$ & $-1.2257\lambda$ \\ \hline
        $\theta > \theta_p$ & $\ell = 0$ & $+1.3932\lambda$ & $-0.0365\lambda$ \\ 
        $45.2^\circ$ & $\ell = 2$ & $+1.3726\lambda$ & $-1.2490\lambda$ \\ \hline
    \end{tabular}
    \caption{Enhancement of IF shift using the LG ($\ell=2$) beam compared to the Gaussian beam ($\ell=0$) for different angles of incidence.}
    \label{tab:placeholder_label}
\end{table}
In both cases ($\ell=0$ and $\ell=2$), the spatial IF receives a significant enhancement, while the spatial GH shift does not change much. This behavior arises because, although a corrugated metallic surface is employed, resulting in complex reflection coefficients \( r_p \) and \( r_s \), and substitution of these values into Eq.~\ref{eq9} yields purely imaginary values of \( Y_{\text{shift}} \). Consequently, the angular IF shift vanishes. 
But the GH shift has both a real and an imaginary part, leading to both the spatial and angular shift (not shown here). That is why, when vortex beams are used, such coupling leads to a significant additional spatial IF shift for both angles of incidence, wherever the additional spatial GH shift is zero. Different spatial GH shift values are obtained for the cases of \( \ell = 0 \) and \( \ell \neq 0 \), which can be attributed to the use of nonparaxial beams rather than strictly paraxial ones.
 In Fig.~\ref{fig3}(e), one can observe some mismatch in the GH shift values near SPR. This discrepancy happens as near the SPR, $r_p$ and $r_s$ are not slowly varying. One has to take higher-order terms in the shift matrix~\cite{r1,weak_IOP} for accuracy. However, in Fig.~\ref{fig3}(f), the IF shift for $\ell=2$ gets enhanced because of the coupling and shows a resonance shape.
\begin{figure}[ht!]
    \centering
    \includegraphics[width= \linewidth]{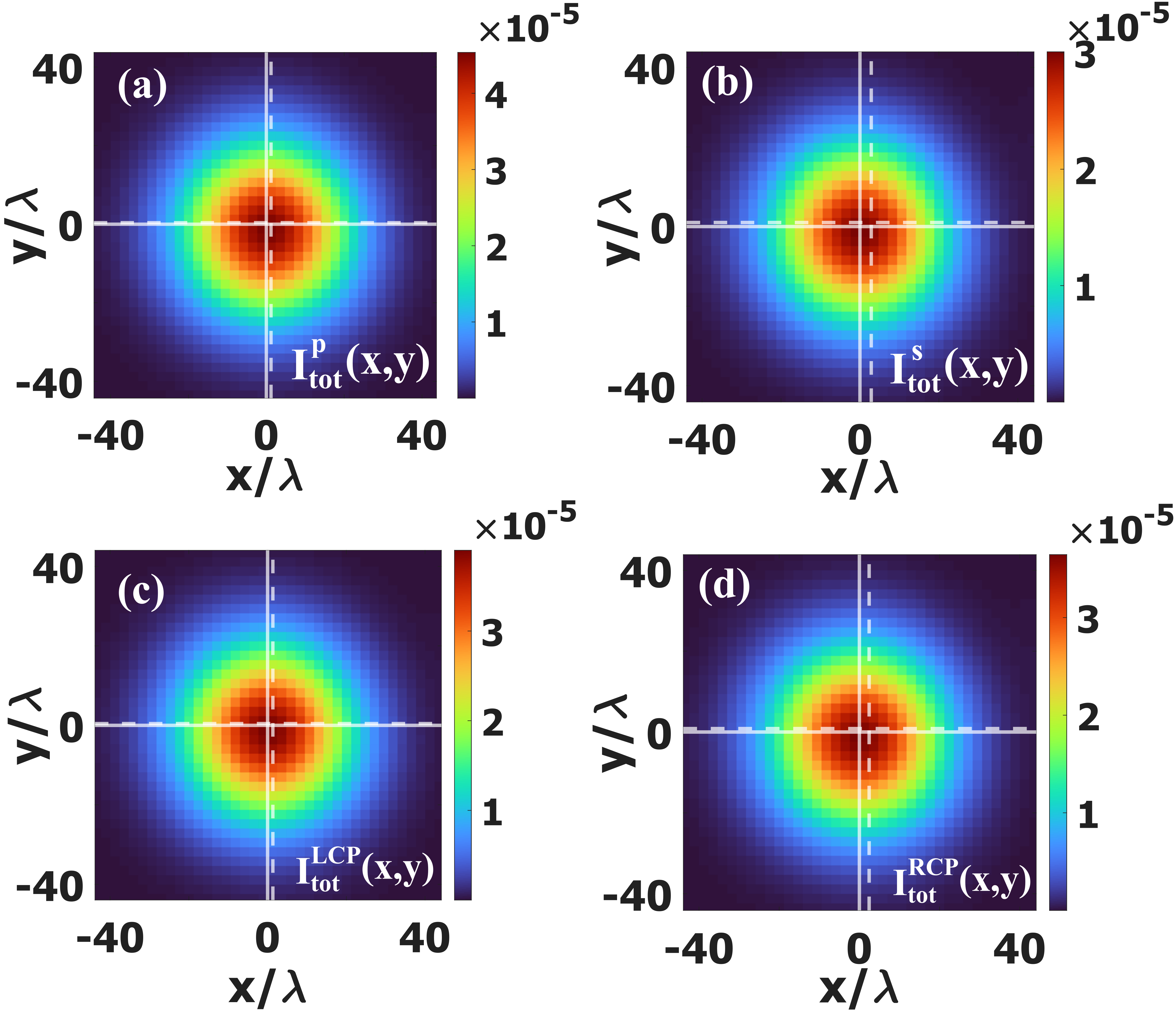}
    \caption{Intensity profiles of the reflected beams for the incident (a) \textit{p}-polarized, (b) \textit{s}-polarized, (c) left circularly polarized, and (d) right circularly polarized Gaussian beams. In all the cases, the angle of incidence is $47.7^\circ$ and the azimuthal angle $\Phi=35^\circ$ (conical diffraction). The dashed and solid crosshairs correspond to the centroids of the reflected and incident beams, respectively.}
    \label{fig4}
\end{figure} 


\subsection{Goos–Hänchen and Imbert–Fedorov shifts in conical diffraction geometry}\label{sec:conical}
Let us now consider the beam shift effects in the case of conical diffraction geometry. In this work, two key effects were demonstrated: first, a pronounced enhancement of the GH shift for both \textit{p}- and \textit{s}-polarized Gaussian beams, and second, an annihilation of the spin Hall effect. To understand this, one has to introduce the off-diagonal terms into the Fresnel Jones matrix due to \textit{p} to \textit{s} and \textit{s} to \textit{p} mode conversion (Figs.~\ref{fig1}(g) and \ref{fig1}(h)) ~\cite{r5}:
\begin{equation}
 F_{c}=\begin{pmatrix}
        r_p  & r_{s\rightarrow p} \\
        r_{p\rightarrow s} & r_s 
    \end{pmatrix}
\label{eq12}
\end{equation}
 The pronounced enhancement of the GH shift near the SPR condition (Figs.~\ref{fig4}(a) and \ref{fig4}(b)) for both \textit{p}- and \textit{s}-polarized Gaussian beams arises from a subtle yet crucial polarization-coupling mechanism at the metal–dielectric interface. In particular, the \textit{s} to \textit{p} polarization conversion channel plays a pivotal role, as it enables the coupling of incident \textit{s}-polarized light to surface plasmon modes—an excitation pathway that is otherwise forbidden within the conventional in-plane framework. This cross-polarization conversion stems from the anisotropic electromagnetic boundary response and the associated SOI of light at the structured plasmonic interface. As a result, even \textit{s}-polarized illumination can indirectly access the plasmon resonance. These effects collectively manifest as a substantial amplification of the spatial GH displacement, thereby revealing the intricate interplay between polarization conversion, plasmonic field confinement, and beam shift phenomena.
Also, because of the off-diagonal terms in Eq.~\ref{eq12}, both linearly polarized Gaussian beams gave significant spatial IF shifts without doing any projections.

Another fascinating phenomenon is the annihilation of the spin-Hall effect. Fig.~\ref{fig4}(c) and~\ref{fig4}(d) represent the intensity profiles of the reflected LCP and RCP Gaussian beams incident conically, where the spatial IF shifts are $+0.4464\lambda$ and $+0.8505\lambda$, respectively, which are not equal and opposite to each other as in the previous case (see Fig.~\ref{fig2} c,f). This is due to the interplay between geometric (spin-dependent) phase and plasmon-induced dynamical phase contributions at the metal–dielectric interface. The off-diagonal coupling terms in Eq.~\ref{eq12} redistribute the spin-dependent transverse momentum, thereby diminishing the usual SOI pathway responsible for opposite lateral shifts. Consequently, the transverse displacement of the LCP and RCP beams no longer remains equal and opposite to each other, resulting in the collapse of the characteristic spin Hall separation. 
\section{CONCLUSIONS}\label{sec:conclusion}

In conclusion, we have presented a comprehensive theoretical and numerical study of the SOI at a resonant corrugated metallic interface and examined the role of surface plasmon resonance in beam-shift phenomena under various structured-beam illumination and detection conditions. To this end, we have employed a vectorial angular spectrum formalism for simulating the diffracted field distributions of different orders. Our particular emphasis was on the GH and IF shifts of the specularly reflected beams, where pronounced enhancements of the GH shifts are observed in the vicinity of the SPR. These originate from the strong phase dispersion of the reflection coefficients and the field enhancements associated with plasmonic excitation. We have also implemented two distinct strategies to amplify the IF shift near the SPR. First, we performed a post-selection onto the $\pm45^\circ$ linear polarization basis, enabling constructive interference of spin-dependent scattering pathways. Second, we employed a weak-measurement protocol by selecting nearly orthogonal pre- and post-polarization states, resulting in weak-value amplification of the subtle transverse beam displacement. The coupling between the orbital angular momentum and polarization-dependent beam shifts for incident optical vortex beams is examined -- before, near, and after the plasmon resonance—highlighting the interplay between spatial and angular degrees of freedom and revealing the dependence of these shifts on both the topological charges and the spin states.

We subsequently extended our analysis to the conical diffraction geometry, where the cross-polarization coupling has induced several unconventional SOI-related behaviors such as annihilation of the conventional spin Hall effect of light, arising from the redistribution of transverse spin-momentum contributions mediated by plasmonic mode coupling. This finding underscores the critical role of interface geometry, mode hybridization, and polarization mixing in shaping spin-orbit-induced optical phenomena. Interestingly, plasmon resonance is found to be excited by an $s$-polarized beam through $s$-to-$p$ polarization conversion, unlike the conventional excitation observed with $p$-polarized illumination.

Overall, our results deepen the understanding of SOI-driven beam shifts at structured plasmonic interfaces and provide a pathway for analyzing more complicated 2D plasmonic metasurfaces. They also provide a recipe for amplifying beam shifts -- that are typically of very small magnitude -- through suitable polarization post-selection schemes including a weak measurement protocol, thus demonstrating an interesting synergy of SOI and weak measurements towards enhancing observable phenomena. These insights hold meaningful implications for emerging applications in nano-photonics, surface-plasmon-based sensing, optical information processing, quantum weak-measurement-enabled precision metrology, and next-generation integrated photonic platforms. 

\section*{Acknowledgement} 
The authors are grateful for the support of the Indian Institute of Science Education and Research Kolkata (IISER Kolkata) and the Ministry of Education, Government of India. One of the authors, S.D.G., is thankful to the Tata Institute of Fundamental Research (Hyderabad) and IISER Kolkata for supporting his honorary positions.

\section*{Appendix A: FORMULATION TO FIND THE REFLECTION COEFFICIENTS}

In this formalism, the Rayleigh expansion method, or equivalently the Floquet expansion method, is followed, and the continuity of the surface field components is applied across the corrugated surface, yielding an infinite and exact set of coupled amplitude equations. All the spatial harmonics, including the evanescent and propagating ones~\cite{r5,r5.1}, are considered.

Let's consider a plane monochromatic incident at an angle $\vartheta$ and with an azimuthal angle $\Phi$ with respect to the grating vector $K$ (Fig. \ref{fig1} (b)). Next, the field distributions corresponding to the incident $p$ and $s$-polarizations in the regions $z > a \sin(Kx)$ (dielectric) and $z < a \sin(Kx)$ (metal) are determined. 

In the $z > a \sin(Kx)$ region, the total field corresponding to the $p$-polarization is:
\begin{equation}
E_{z1} = e^{i\alpha_{0}y} 
\left[ 
A_{0}e^{i(\gamma_0x-\beta_0z)} 
+ 
\sum_{n} B_{n}e^{i(\beta_nz+\gamma_nx)} 
\right]
\tag{A1}
\end{equation}
and for the $s$-polarization is:
\begin{equation}
H_{z1} = e^{i\alpha_{0}y} 
\left[ 
\tilde{A}_{0}e^{i(\gamma_0x-\beta_0z)} 
+ 
\sum_{n} \tilde{B}_{n}e^{i(\beta_nz+\gamma_nx)} 
\right]
\tag{A2}
\end{equation}
Similarly, for the $z < a \sin(Kx)$, the field corresponding to the $p$-polarization is:
 \begin{equation}
E_{z2} = e^{i\alpha_{0}y} 
\left[  
\sum_{n} C_{n}e^{i(\gamma_nx-\tilde{\beta}_nz)} 
\right]
\tag{A3}
\end{equation}
and for the $s$-polarization:
\begin{equation}
H_{z2} = e^{i\alpha_{0}y} 
\left[  
\sum_{n} \tilde{C}_{n}e^{i(\gamma_nx-\tilde{\beta}_nz)} 
\right]
\tag{A4}
\end{equation}
where 
\begin{align*}
\alpha_{0} &= k_{0}\sqrt{\epsilon_{d}} \sin\vartheta \sin\Phi, \notag\\
\gamma_{0} &= k_{0}\sqrt{\epsilon_{d}} \sin\vartheta \cos\Phi, \notag\\
\beta_{0}^{2} &= k_{0}^{2}\epsilon_{d} - \gamma_{0}^{2} - \alpha_{0}^{2}, \notag\\
\gamma_{n} &= \gamma_{0} + nK, \quad n = 0, \pm1, \pm2, \ldots, \notag\\
\beta_{n}^{2} &= k_{0}^{2}\epsilon_{d} - \gamma_{n}^{2} - \alpha_{0}^{2}, \notag\\
\tilde{\beta}_{n}^{2} &= k_{0}^{2}\epsilon_m - \gamma_{n}^{2} - \alpha_{0}^{2}.
\end{align*}
Here, $A_0$, $\tilde{A}_0$ are the incident amplitudes, $B_n$, $\tilde{B}_n$ are the reflected spatial-harmonic amplitudes, and $C_n$ ,$\tilde{C}_n$ are transmitted spatial-harmonic amplitudes. To assign a clear physical interpretation of the different amplitude components~\cite{r5}, we define
\begin{equation}
 A_0=Q\sin(\vartheta)/\epsilon_d
 \tag{A5}
\end{equation}
\begin{equation}
 \tilde{A_0}=\tilde{Q}\sin(\vartheta)\epsilon_d
 \tag{A6}
\end{equation}
For a \textit{p}-polarized incident wave, we set $Q = 1, \tilde{Q} = 0$ and for \textit{s}-polarized wave $Q = 0, \tilde{Q} = 1$. 

Next, the boundary conditions are applied to the corrugated surface.
\begin{equation}
E_{y1}\big|_{z = a \sin Kx} = E_{y2}\big|_{z = a \sin Kx}
\tag{A7}
\end{equation}
\begin{equation}
H_{y1}\big|_{z = a \sin Kx} = H_{y2}\big|_{z = a \sin Kx}
\tag{A8}
\end{equation}
\begin{equation}
\left( E_{x1} t_{x} + E_{z1} t_{z} \right)\bigg|_{z = a \sin Kx} = 
\left( E_{x2} t_{x} + E_{z2} t_{z} \right)\bigg|_{z = a \sin Kx}
\tag{A9}
\end{equation}
\begin{equation}
\left( H_{x1} t_{x} + H_{z1} t_{z} \right)\bigg|_{z = a \sin Kx} = 
\left( H_{x2} t_{x} + H_{z2} t_{z} \right)\bigg|_{z = a \sin Kx}
\tag{A10}
\end{equation}
Here, $t_x$ and $t_z$ are the $x$ and $z$ components of the unit vector
$\boldsymbol{t}$, tangential to the surface:
\begin{equation}
    \boldsymbol{t} = \left\{ 1 + \left[ aK \cos(Kx) \right]^2 \right\}^{-1/2} (1, aK \cos(Kx), 0)
    \label{eq:placeholder_label}
\end{equation}
The Bessel function expansions are used to evaluate the following exponents:
\begin{equation}
e^{i\beta_{n} a \sin Kx} 
= 
\sum_{m=-\infty}^{\infty} 
e^{imKx} J_{m}(\beta_{n} a)
\tag{A12}
\end{equation}
and we obtain the infinite set of amplitude equations. In the final step, the infinite system was truncated to a finite order, and the solutions were computed for this truncated system.
\bibliographystyle{apsrev4-2}
\bibliography{apsamp}
\end{document}